\documentclass[12pt]{iopart}
\usepackage{iopams}
\usepackage{epsfig}
\bibstyle{plain}
\begin{document}

\title{Coevolution of teaching activity promotes cooperation}

\author{Attila Szolnoki$^1$ and Matja{\v z} Perc$^2$}
\address{$^1$Research Institute for Technical Physics and Materials Science, P.O. Box 49, H-1525 Budapest, Hungary\\ $^2$Department of Physics, Faculty of Natural Sciences and Mathematics, University of Maribor, Koro{\v s}ka cesta 160, SI-2000 Maribor, Slovenia}
\ead{szolnoki@mfa.kfki.hu, matjaz.perc@uni-mb.si}

\begin{abstract}
Evolutionary games are studied where the teaching activity of players can evolve in time. Initially all players following either the cooperative or defecting strategy are distributed on a square lattice. The rate of strategy adoption is determined by the payoff difference and a teaching activity characterizing the donor's capability to enforce its strategy on the opponent. Each successful strategy adoption process is accompanied with an increase in the donor's teaching activity. By applying an optimum value of the increment this simple mechanism spontaneously creates relevant inhomogeneities in the teaching activities that support the maintenance of cooperation for both the prisoner's dilemma and the snowdrift game.
\end{abstract}

\maketitle

\section{Introduction}

Cooperation amongst selfish individuals is an essential underpinning of modern human societies and wildlife coexistence alike. Revelation of mechanisms supporting cooperation against the fundamental principles of Darwinian selection is therefore of key interest within many branches of social and natural sciences \cite{axelrod_84}. Although verbal arguments to address the issue abound, the puzzle of how and why individuals overcome selfishness in order to subdue their actions to the common good presents a formidable challenge within the scientific community. A common mathematical framework of choice for addressing the many subtleties of cooperation within groups of selfish individuals is the evolutionary game theory \cite{ms_82, hofbauer_88, weib_95, gintis_00, nowak_06}, and in particular the prisoner's dilemma as well as the snowdrift game have become widely adopted for this purpose. In both games the mutual cooperation warrants the highest collective payoff that is equally shared amongst the players. Mutual cooperation is, however, challenged by the defecting strategy that promises the defector a higher income at the expenses of a neighboring cooperator. The crucial difference between these two games is the way in which defectors are punished if faced one another. In the prisoner's dilemma game a defector encountering another defector still earns more than a cooperator facing a defector, whilst in the snowdrift game the ranking of these two payoffs is reversed. Thus, in the snowdrift game a cooperator facing a defector receives a higher payoff than a defector playing with another defector. This seemingly minute difference between both games can have a rather profound effect on the evolutionary success of the two strategies. Particularly for the spatial version of both games, it has been reported that while by the prisoner's dilemma nearest neighbor interactions generally facilitate cooperation \cite{nowak_n92b} this is often not the case by the snowdrift game \cite{hauert_04}. In contrast, the facilitative effect of the scale-free topology to promote cooperation prevails in both \cite{santos_prl05}. Given the difficulties associated with payoff rankings in experimental and field work \cite{milinski_prsb97, turner_n99}, the two games certainly deserve separate attention and have rightfully acquired a central role within the pursue of cooperation in evolutionary game theory.

The seminal works of Nowak and May \cite{nowak_n92b, nowak_ijbc93} spawned many studies and new approaches aimed towards resolving the dilemma \cite{huberman_93, lindgren_94, wedekind_00, hauert_02, kerr_02, jimenez_08}, whereby perhaps the most direct extension came in the form of evolutionary games on complex networks \cite{abramson_01, ebel_pre02, holme_03, wu_05, tomassini_05, vukov_05, szabo_06, wang_06, poncela_njp07, wang_07, rong_07, chen_08} that were comprehensively reviewed in \cite{szabo_cm06}. Within latter it has become apparent that heterogeneities amongst players, especially in the form of scale-free degree distribution warranted by namesake networks \cite{albert_02, boccaletti_06}, can have strong facilitative effects on the evolution of cooperation \cite{santos_prl05, santos_prb05}. Subsequently, the positive impact of heterogeneity, albeit of a different origin ({\it i.e.} not related to the structure of host networks), has been confirmed also via the introduction of noise to the payoffs \cite{perc_06a, tanimoto_07}, inhomogeneities by strategy adoption probabilities \cite{kim_02, wu_06, szolnoki_epl07}, social diversity \cite{perc_pre08}, as well as bimatrix games \cite{fort_08}. However, besides offering new ways to sustain cooperation, these mechanisms also pose new puzzles that need to be addressed; like how do such heterogeneities come about, do they evolve and if yes in what way, and what seem to be their most likely origins. Pacheco {\it et. al.} have recently made important steps in this direction by extending the subject of games on graphs via active or dynamical linking \cite{pacheco_jtb06, pacheco_prl06}, showing that the latter may help to maintain cooperative behavior, whereas somewhat earlier studies employing random or intentional rewiring procedures \cite{ebel_02, zimmermann_04, zimmermann_05, perc_06b} came to similar conclusions. Moreover, recent studies separately addressing interaction and strategy adoption graphs \cite{ohtsuki_prl07, ohtsuki_jtb07, wu_pre07} also contributed substantially to revealing mechanisms behind the survival and promotion of cooperation within the prisoner's dilemma and the snowdrift game, in particularly showing that the separation of the two graphs completely disables the survival of cooperators if the overlap between them is zero.

Inhomogeneities amongst members of human and animal societies are common. Indeed, many phrases and titles have been invented to distinguish influential individuals from those having little impact, and more often than not `being influential' is reserved for few selected individuals only. Previous studies already highlighted that such differences between players may beneficially serve common interests \cite{szolnoki_epl07, guan_pre07}. The scope of this paper is to investigate how such heterogeneities develop by studying the coevolution of teaching activity (or influence briefly) and strategy within the spatial prisoner's dilemma and snowdrift game. In both games the so-called influence of each individual is the quantity determining heterogeneity of participating players, specifically affecting the ability of each to enforce its strategy on the opponent, whereby in accordance with logical reasoning, influential individuals are much more likely to reproduce than players with low influence. We find that, although initially all players have the same influence, the employed rule for the evolution of influence quickly results in a heterogeneous distribution of the latter, which in turn facilitates the evolution of cooperation in accordance with the established reasoning concerning the impact of heterogeneities amongst players outlined for example in \cite{santos_prl05, perc_pre08}. Despite the simplicity of the employed coevolution rule for influence, our model accounts for the often-observed large segregation in real life based exclusively on the theoretical framework of evolutionary game theory, and moreover, shows that the resulting exponential distribution of influence, emerging spontaneously from an initially unpreferential setup, provides permanent support for the cooperative strategy in the prisoner's dilemma as well as the snowdrift game. Our results convey the potentially disturbing message that large differences in status may arise spontaneously, and although they might evoke discomfort within the majority that is disprivileged, they are vital for keeping the population in a cooperative state, especially so if temptations to defect are large.

The remainder of this paper is organized as follows. In the next section we describe the two employed evolutionary games and the protocol for the coevolution of influence. Section 3 is devoted to the presentation of results, whereas in the last section we summarize and discuss their implications.

\section{Game definitions and setup}

As already noted, we use the spatial prisoner's dilemma and snowdrift game for the purpose of this study. In accordance with common practice, the prisoner's dilemma is characterized by the temptation $T = b$, reward $R = 1$, and both punishment $P$ as well as the suckers payoff $S$ equaling $0$, whereby $1 < b \leq 2$ ensures a proper payoff ranking. The snowdrift game, on the other hand, has $T = \beta$, $R = \beta - \frac{1}{2}$, $S = \beta - 1$ and $P = 0$, whereby $r = \frac{1}{2 \beta - 1}$ remaining within the unit interval ensures that $T > R > S > P$. To eschew effects of complex host graph topologies, we employ a regular $L \times L$ square lattice with periodic boundary conditions irrespective of which game applies. Initially, a player on the site $x$ is designated as a cooperator ($s_x=C$) or defector ($D$) with equal probability, and the game is iterated in accordance with the Monte Carlo (MC) simulation procedure comprising the following elementary steps. First, a randomly selected player $x$ acquires its payoff $P_x$ by playing the game with its four nearest neighbors. Next, one randomly chosen neighbor, denoted by $y$, also acquires its payoff $P_y$ by playing the game with its four neighbors. Last, player $x$ tries to enforce its strategy $s_x$ on player $y$ in accordance with the probability
\begin{equation}
W(s_y \rightarrow s_x)=w_x \frac{1}{1+\exp[(P_y-P_x)/K]},
\label{eq:prob}
\end{equation}
where $K$ denotes the amplitude of noise and $w_x$ characterizes the strength of influence (or teaching activity) of player $x$. Importantly, $w_x$ is also subjected to an evolutionary process in accordance with the following protocol that applies to both games alike. Initially, all players are given the minimal influence factor $w_x = w_m \ll 1$, thus assuring a nonpreferential setup of the game. Note, however, that $w_m$ must be positive in order to avoid frozen states, and hence we use $w_m = 0.01$ throughout this study. Next, every time player $x$ succeeds in enforcing its strategy on $y$ the influence $w_x$ is increased by a constant positive value $\Delta w \ll 1$ according to $w_x \rightarrow w_x + \Delta w$. Finally, the evolution of influence is stopped for all players as soon as one $w_x$ reaches $1$. Despite being strikingly simple and relatively fast to finish (typically around $100$ MC steps), the proposed protocol for the coevolution of influence is remarkably robust, delivering conclusive results with respect to the final distribution of $w_x$. It is worth emphasizing that the evolution of influence takes place on a much faster time scale than the simultaneous evolution of the strategy distribution. Nevertheless, we would like to note that the extremely short evolution time for $w_x$ suffices completely for a robust establishment of the stationary distribution of $w_x$. We have verified this by employing an alternative evolution protocol by which $w_x$ was allowed to grow also past $1$, only that then $w_x$ was normalized according to $w_x \rightarrow \frac{w_x}{w_{max}}$ ($w_{max}>1$ being the maximal out of all $w_x$ at any given instance of the game) to assure that the teaching activity remained bounded to the unit interval. This alternative rule for the evolution of influence yields identical results with respect to stationary fractions of strategies as well as distributions of $w_x$ as the halted version used throughout this study, which directly implies that `who gets to run the show' ({\it i.e.} who has the largest $w_x$ assigned) is determined already at the very infancy of the game.

MC results presented below were obtained on populations comprising $400 \times 400$ to $1600 \times 1600$ individuals, whereby the stationary fraction of cooperators $\rho_C$ was determined within $5\cdot10^5$ to $3\cdot10^6$ MC steps after sufficiently long transients were discarded. Moreover, due to the much shorter temporal scale characterizing the evolution of influence and its resulting highly heterogeneous distribution, final results were additionally averaged over $30$ to $300$ independent runs for each set of parameter values in order to assure accuracy. Noteworthy, due to the resulting heterogeneous distribution of influence the current coevolutionary model demands similar computational resources as robust simulations of evolutionary games on scale-free networks, where several independent runs by the same parameters are also necessary to take into account the stochastic feature of the host graph topology.

\section{Results}

\begin{figure}
\begin{center} \includegraphics[width = 11cm]{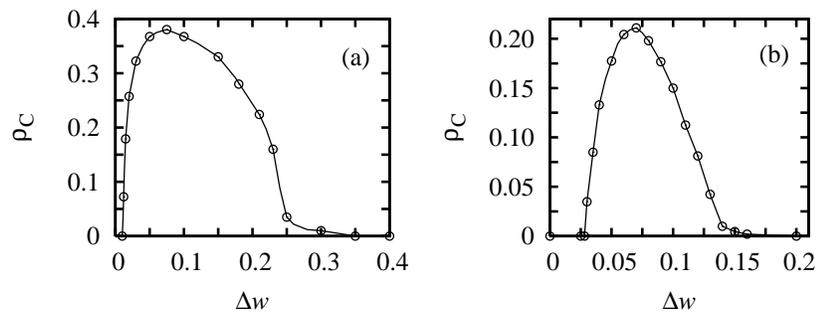}
\caption{\label{fig1} Promotion of cooperation due to the coevolution of influence via $\Delta w$. Both panels show the stationary fraction of cooperators $\rho_C$ in dependence on $\Delta w$, whereby the optimal value for the latter equals $\approx 0.07$ irrespective of which game applies. Panel (a) shows results for the prisoner's dilemma ($b = 1.05$, $K = 0.1$) and (b) for the snowdrift ($r = 0.6$, $K = 2$) game.}
\end{center}
\end{figure}

In what follows, we will systematically analyze effects of different $\Delta w$, $K$ and payoff values on the evolution of cooperation within the two employed games. Throughout this section results for the prisoner's dilemma and the snowdrift game will be shown and commented in a parallel fashion for the purpose of better comparison options.

We start revealing the properties of the above-introduced model by examining the impact of evolving influence $w_x$ on the stationary fraction of cooperators $\rho_C$ within the two employed games. Figure~\ref{fig1} shows results, obtained by a given combination of the temptation to defect (either $b$ or $r$) and strategy adoption uncertainty $K$, separately for the prisoner's dilemma and the snowdrift game in panels (a) and (b), respectively. Evidently, $\Delta w = 0$ corresponds to the traditional version of both games where players are not distinguished, and thus $\rho_C = 0$ due to large $b$ and $r$. For small $\Delta w$ the impact on $\rho_C$ remains marginal because the resulting heterogeneity amongst players is minute, {\it i.e.} influential players fail to differ relevantly from the disprivileged individuals. However, as $\Delta w$ exceeds a threshold value a remarkable increase of $\rho_C$ can be observed, thus indicating that an optimal distribution of influence warranting the most significant benefits for cooperators has been established. By further increasing $\Delta w$ the cooperation-facilitative effect again deteriorates, which is due to the too fast stop of the evolution of influence. In particular, large $\Delta w$ (comparable to $1$) essentially leave the whole population in a homogeneous state characterized by $w_x \approx w_m$, whereas the very few influential players having $w_x$ close to $1$ simply don't generate a noticeable impact on the evolution of the two strategies. Irrespective of payoff rankings differentiating the two games though, the optimal value for the increment of influence equals $\Delta w \approx 0.07$, which we will therefore use also in all subsequent calculations.

\begin{figure}
\begin{center} \includegraphics[width = 11cm]{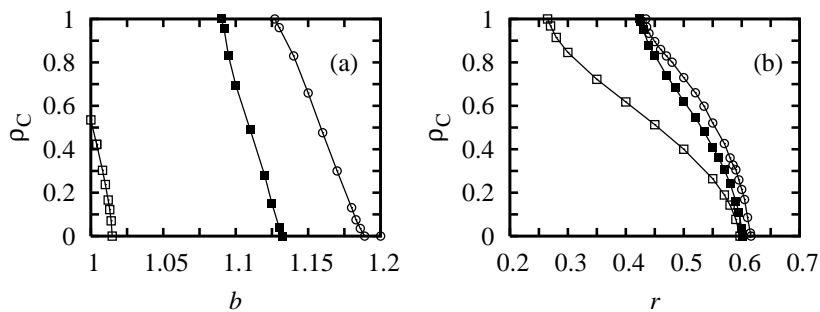}
\caption{\label{fig2} Maintenance of cooperation via solely the spatial structure (open squares), the introduction of fixed influence heterogeneity (filled squares; see text for details), and via the evolution of influence by setting $\Delta w = 0.07$ (open circles). Panels (a) and (b) show results for the prisoner's dilemma and the snowdrift game, respectively. Evidently, the current model, encompassing the evolution of influence, is most successful in promoting the cooperative strategy. All curves were obtained by setting $K = 2$.}
\end{center}
\end{figure}

To make the initial observations regarding the promotion of cooperation via the coevolution of influence more precise, we present $\rho_C$ in dependence on the whole relevant span of $b$ and $r$ for the prisoner's dilemma and the snowdrift game in Fig.~\ref{fig2}. For comparisons, in addition to showing results obtained with the presently introduced setup encompassing the coevolution of influence (open circles), results obtained with the traditional spatial versions of the two games [obtained simply by setting $w_x = 1$ for all $x$ in Eq.~(\ref{eq:prob}); open squares], and by assigning half of the players as having a fixed lower influence $w_x = 0.1$ (filled squares), are shown as well. Note that the latter setup proved to be optimal when fixed heterogeneous teaching activities were applied in a previous study \cite{szolnoki_epl07}. Indeed, the presently introduced model warrants the best facilitation of cooperation in both games, clearly improving the performance of solely spatial interactions and fixed heterogeneity of influence. Nevertheless, while by the prisoner's dilemma game the positive effect on cooperation is evident across the whole relevant span of $b$, individuals engaging in the snowdrift game profit the most from the coevolution of influence by smaller $r$, whereas additional benefits with respect to previous models deteriorate continuously as $r$ increases. Despite this discrepancy brought about by the different payoff rankings of the two employed games, the concluding observation is that the spontaneous coevolution of influence from an initially unpreferential state results in advantageous environment for cooperation compared to models without coevolutionary ingredients.

\begin{figure}
\begin{center} \includegraphics[width = 11cm]{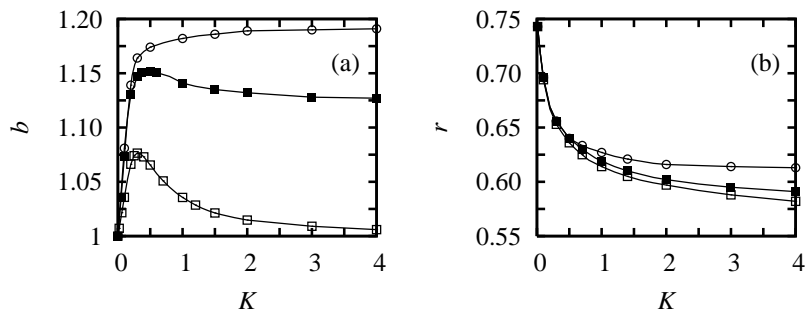}
\caption{\label{fig3} Phase separation lines on the $K - b$ parameter plane for the prisoner's dilemma (a) and the snowdrift (b) game. Lines denote the border separating mixed $C + D$ (below) and pure defector $D$ (above) states. Compared to the other two models (open and filled squares), the evolution of influence (open circles) facilitates cooperation increasingly better as $K$ increases. Employed parameter values are identical to those used in Fig.~\ref{fig2}.}
\end{center}
\end{figure}

To support the above suggested picture, phase separation lines on the $K-b$ parameter plane are presented in Fig.~\ref{fig3} for both games. Symbols show results for the same models as in Fig.~\ref{fig2}, whereby below the lines cooperators and defectors coexist while above a homogeneous defector state always prevails. Notably, these phase transitions exist irrespective of the magnitude of $K$ as long as the latter is finite. Results summarized in Fig.~\ref{fig3} evidence that, as the strategy adoption uncertainty $K$ increases, the evolution of influence is increasingly more successful by promoting cooperation in comparison to previous approaches. In addition however, it can be observed that the overall impact of increasing $K$ is exactly opposite by the two employed games; continuously facilitating cooperation by the prisoner's dilemma game [open circles in panel (a)] on one hand, while monotonously deteriorating it by the snowdrift game [open circles in panel (b)] on the other. Noteworthy, while the impact of different $K$ on the evolution of cooperation in the prisoner's dilemma game has been studied accurately already in \cite{szabo_pre05}, the complete phase diagram for the snowdrift game has not yet been presented. The observed difference of impact imposed by increasing $K$ concurs nicely with previously reported discrepancies attributed to the different payoff rankings of the two games \cite{hauert_04}, and moreover, supports the fact that uncertainties may in general facilitate defection in the snowdrift game \cite{perc_ijbc07}.

\begin{figure}
\begin{center} \includegraphics[width = 11cm]{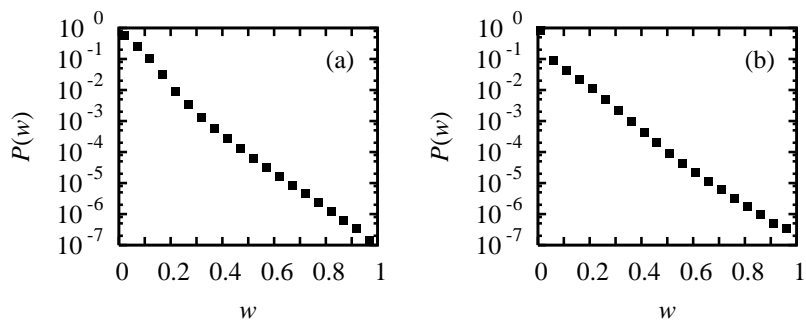}
\caption{\label{fig4} Final distributions of influence $P(w)$ in the two studied games obtained via $\Delta w = 0.07$. Panel (a) shows results for the prisoner's dilemma ($b = 1.05$, $K = 0.1$) and (b) for the snowdrift ($r = 0.6$, $K = 2$) game. Note that in both panels the $y$ axis has a logarithmic scale, and thus the depicted linear dependence of $P(w)$ indicates an exponential distribution that emerges irrespectively of which game applies.}
\end{center}
\end{figure}

Finally, it remains of interest to examine the resulting distributions of influence $P(w)$ emerging within the two employed games. Given the fact that substantial promotion of cooperation was in the past often associated with strongly heterogeneous states, either in form of the host network \cite{santos_prl05} or social diversity \cite{perc_pre08}, it is reasonable to expect that $P(w)$ will exhibit a highly heterogeneous outlay as well. Results presented in Fig.~\ref{fig4} clearly attest to this expectation as symbols in both panels can be well approximated with a straight line in a semi-log scale, hence indicating an exponential distribution of $w_x$. These distributions are rather robust and independent of $K$ or payoff values. The highly heterogeneous final state is crucial for the fortified facilitative effect on cooperation outlined in Figs.~\ref{fig2} and \ref{fig3}, in particular since it incubates cooperative clusters around individuals with high $w_x$. On the contrary, since the positive feedback of imitating environment is not associated with influential defectors they therefore fail to survive even if temptations to defect are large. As already noted, a similar behavior underlies the cooperation-facilitating mechanism reported for the scale-free network where players with the largest connectivity (presently equivalent to those having $w_x$ close to 1) also act as robust sources of cooperation in the prisoner's dilemma and the snowdrift game \cite{santos_prl05}. The presently reported spontaneous emergence of the heterogeneous distribution of influence from an initially unpreferential state within the framework of evolutionary game theory suggests that even very simple coevolutionary rules might lead to strong segregations amongst participating players, which are arguably advantageous for flourishing cooperative states. Noteworthy, the beneficial impact of segregated players has already been reported in a previous study where, however, this constraint was artificially postulated \cite{perc_pre08}.

\section{Summary}

We have studied the coevolution of influence and strategy in the spatial prisoner's dilemma and snowdrift game. We show that a highly inhomogeneous distribution of influence may emerge spontaneously from an initially completely unpreferential setup, thus providing insights that shed light on possible origins of heterogeneity within the framework of evolutionary game theory. Given the simplicity of the newly introduced rule for the coevolution of influence, we believe that the present approach can be extended further to account also for other forms of heterogeneity that may be associated with individuals indulging into evolutionary games. Noteworthy, a conceptually similar approach has been adopted recently by Garlaschelli \textit{et al.} \cite{garlaschelli_07}, who studied the interplay between topology and dynamics on a model in which the network was shaped by a dynamical variable of the Bak-Sneppen evolution model, and there also a highly heterogeneous state could emerge spontaneously above a certain threshold. An interesting alternative model for studying coevolutionary aspects by social dilemmas has been presented in \cite{jesus_ax} as well. Moreover, we reveal that the spontaneously emerging heterogeneous distribution of influence warrants the most potent promotion of cooperation in both studied games, which supports the notion that the coevolution of a secondary quantity affecting the distribution of strategies might yield excessive benefits for the cooperative trait substantially surpassing those that can be expect from a manually introduced bi-heterogeneous state or spatiality alone. Resulting heterogeneous distributions of influence also demonstrate the strongly asymmetric flow of strategy adaptations between players, which was already observed when the scale-free host topology was applied \cite{szolnoki_pa08}.

In sum, presented results confirm that the presence of influential leaders is advantageous for cooperation, and that it may emerge spontaneously even under simple coevolutionary rules. Hence, the large segregation of individuals observed in many human and animal societies is vital for the sustainability of cooperation, and it seems just to ask of the less-fortunate to accept such social states, but of course only so far as the leaders themselves subdue to the cooperative trait.

\ack
We thank Gy\"{o}rgy Szab\'{o} for valuable discussions and useful comments. The authors acknowledge support from the Hungarian National Research Fund (grant K-73449) and the Slovenian Research Agency (grant Z1-9629).

\end{document}